\documentclass[aps,prb,twocolumn,superscriptaddress,showkeys,amsmath,amssymb]{revtex4-1}
\usepackage{amsmath,booktabs,multirow}
\usepackage{hyperref}
\usepackage{graphicx}
\AtBeginDocument{
\heavyrulewidth=.08em
\lightrulewidth=.05em
\cmidrulewidth=.03em
\belowrulesep=.65ex
\belowbottomsep=0pt
\aboverulesep=.4ex
\abovetopsep=0pt
\cmidrulesep=\doublerulesep
\cmidrulekern=.5em
\defaultaddspace=.5em
}

\begin{document}

\title[H on Sb(111) measured by HAS]{Adhesion Properties of Hydrogen on Sb(111) Probed by Helium Atom Scattering}

\author{Patrick Kraus}
\author{Christian G\"{o}sweiner}
\author{Anton Tamt\"{o}gl}
\author{Florian Apolloner}
\author{Wolfgang E. Ernst}
\affiliation{Institute of Experimental Physics, Graz University of Technology - Petersgasse 16, 8010 Graz, Austria}


\begin{abstract}
We have carried out a series of helium atom scattering measurements in order to characterise the adsorption properties of hydrogen on antimony(111). Molecular hydrogen does not adsorb at temperatures above 110 K in contrast to pre-dissociated atomic hydrogen. Depending on the substrate temperature, two different adlayer phases of atomic hydrogen on Sb(111) occur. At low substrate temperatures ($110~$K), the deposited hydrogen layer does not show any ordering while we observe a perfectly ordered $(1\times 1)$ H/Sb(111) structure for deposition at  room temperature. Furthermore, the amorphous hydrogen layer deposited at low temperature forms an ordered overlayer upon heating the crystal to room temperature. Hydrogen starts to desorb at $T_m = 430~$K which corresponds to a desorption energy of $E_{des}=(1.33\pm0.06)~$eV. Using measurements of the helium reflectivity during hydrogen exposure at different surface temperatures, we conclude that the initial sticking coefficient of atomic hydrogen on Sb(111) decreases with increasing surface temperature. Furthermore, the scattering cross section for the diffuse scattering of helium from hydrogen on Sb(111) is determined as $\Sigma = (12 \pm 1)~\mbox{\AA}^{2}$.
\end{abstract}

\maketitle

\section{Introduction}
\label{sec:introduction}
The semimetal antimony (Sb), one of the essential components in the newly discovered group of topological insulators\cite{Zhang2009}, has gained wide attention recently. While antimony itself is a topological semimetal\cite{Hsieh2009}, Sb nanofilms are proposed to be topological insulators\cite{Zhang2012} and are often stated as interesting candidates for applications in spintronics\cite{Bian2012}. Recently, first-principles calculations showed that the adsorption of hydrogen (H) on antimony thin films is capable of modulating the topological surface states: Adsorption of H on Sb induces a huge band gap and a quantum phase transition from an ordinary insulator to a non-trivial topological phase\cite{Wang2015}.\\
The adsorption of hydrogen on metal surfaces has been the subject of many experimental and theoretical efforts in the last decades. Hydrogen, by virtue of its single electron, provides also an ideal system for modelling gas-surface interactions\cite{Christmann1995,Christmann1988}. However, very little experimental data exists for the adsorption properties of semimetal surfaces apart from graphene\cite{Lin2015} and to our knowledge there are no experimental studies upon the adsorption of atomic hydrogen on Sb(111).\\
It has been shown via scanning tunneling microscopy (STM)\cite{Stegemann2004}, low electron energy diffraction (LEED)\cite{Jona1967,Bengio2007} and helium atom scattering (HAS)\cite{Mayrhofer2013,Mayrhofer2013b}, that neither reconstructions nor mentionable relaxations occur at the (111) surface of Sb. Moreover, it has been claimed that antimony is chemically highly inert\cite{Shan2011}, with the only exception of surface oxidation occurring at elevated temperatures $>260~^{\circ}\mbox{C}$\cite{Rosenberg1960}.\\
We present helium atom scattering measurements which allow us to characterise the adsorption behaviour of Sb(111) with respect to atomic and molecular hydrogen. Neutral He atom beams with energies of typically 10-20~meV are perfectly suited to probe this system in an inert, completely non-destructive manner\cite{Farias1998}. HAS provides an accurate description of the surface charge density corrugation as seen by He atoms at thermal energies\cite{Tamtoegl2015,Kraus2015} whereas e.g. electron scattering is not well suited to detect hydrogen overlayers due to the very small cross-section of hydrogen atoms for electrons\cite{Woell2004}.

\section{Experiment}
\label{sec:experiment}
All measurements presented in this manuscript have been conducted on the local helium atom scattering apparatus\cite{Tamtoegl2010}. An energetically narrow ($< 2\%~E_0$) helium beam is produced via supersonic expansion from a high pressure chamber (50~bar) into the source chamber vacuum ($\sim 10^{-7}$~mbar). The central portion of the spatial gas distribution is selected using a conical skimmer at a distance of approximately 10~mm from the nozzle. In the main measuring chamber of the apparatus, the sample can be cleaned by $\mathrm{Ar^{+}}$-ion sputtering and annealing.\\
The surface can be modified by hydrogen exposure using either thermal molecular hydrogen or an atomic hydrogen source. The source produces a beam of atomic hydrogen from molecular hydrogen which flows through a heated tungsten capillary where it dissociates (see \cite{Eibl1998} for a characterisation of the source). The main chamber additionally features a quadrupole mass analyser for accurate residual gas analysis and a combined LEED and Auger-electron-spectroscopy (AES) apparatus for surface analysis.\\
The initial scattering angle of the helium on the so-prepared surface can be controlled by mounting the sample on a seven-axis manipulator, centred in the main chamber of the apparatus. The angular intensity of scattered helium atoms can be detected along a fixed total scattering angle using a quadrupole mass analyser.

\subsection{Sb(111) substrate}
\label{sec:substrate}
The substrate used for the hydrogen adsorption in this study was an Sb(111) single crystal. The properties of the clean Sb(111) crystal surface have thoroughly been investigated by our group using HAS and Close-Coupling (CC) simulations\cite{Tamtoegl2013,Mayrhofer2013,Mayrhofer2013b,Kraus2014}. Bulk antimony exhibits the rhombohedral A7 structure, with puckered bilayers perpendicular to the $[111]$ direction. The resulting (111) surface can be described using a hexagonal unit cell with a lattice constant of $a~=~(4.3084~\pm~0.0002)~\mathrm{\AA}$\cite{Bengio2007,Stegemann2004}. The surface has two high-symmetry directions, $\overline{\Gamma \mathrm{M}}$ and $\overline{\Gamma \mathrm{K}}$. All HAS measurements within this study were performed with the scattering direction aligned parallel to the $\overline{\Gamma \mathrm{M}}$-direction. As in the present study, all HAS experiments were carried out keeping the clean sample at a base pressure of $10^{-10}$ mbar to avoid the adsorption of residual gas atoms.

\section{Results}
\label{sec:results}

\subsection{Molecular Hydrogen Adsorption}
\begin{figure}[htb]
\includegraphics[width=0.8\columnwidth]{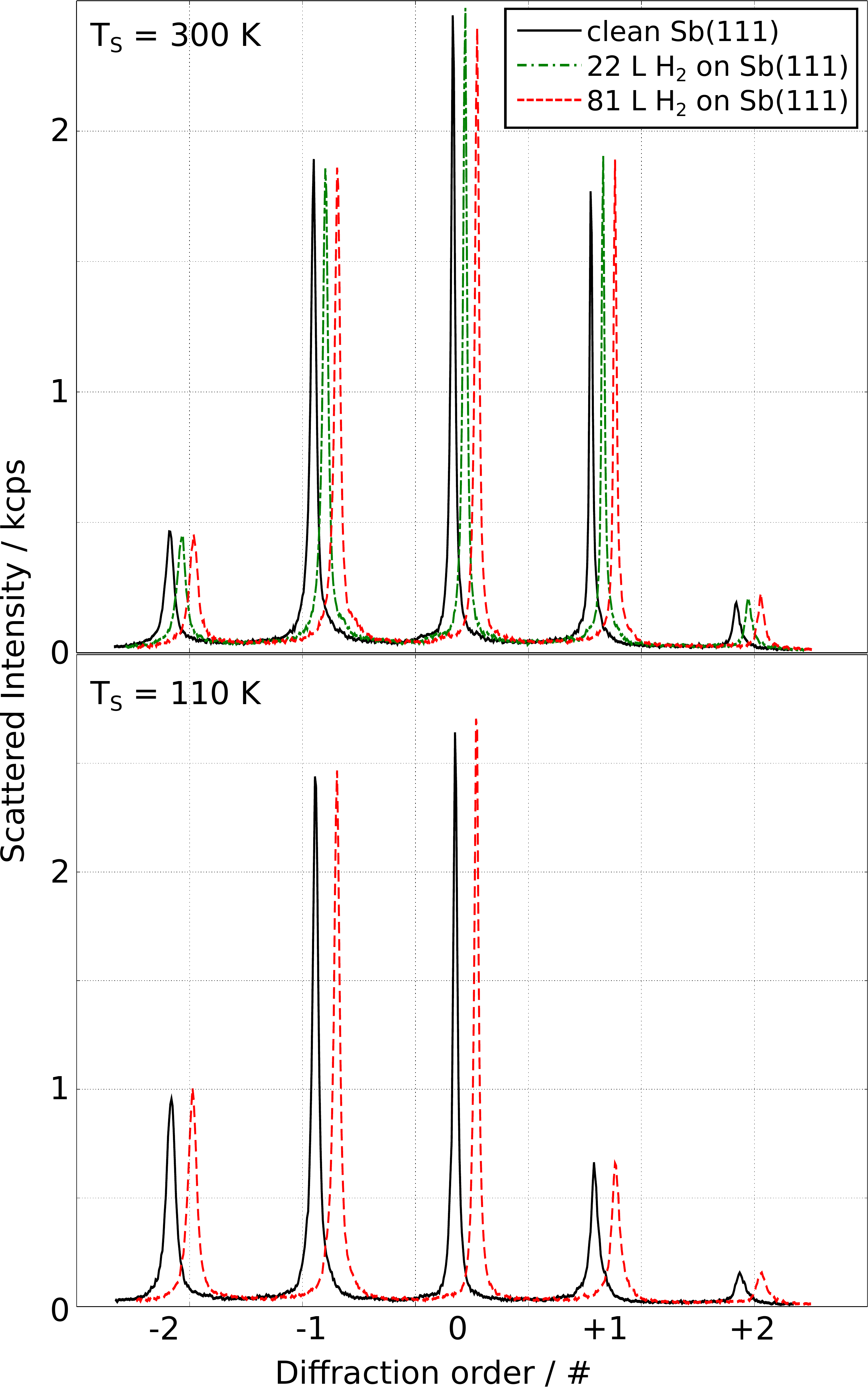}
\caption{Angular scattering spectra of the clean Sb(111) surface and after exposure to H$_2$ along the $\overline{\Gamma \mathrm{M}}$ direction and at two surface temperatures. The surface seems largely inert to application of even large amounts of molecular hydrogen. The recorded spectra have been shifted by several degrees to facilitate the presentation. The asymmetry in the 110~K measurement originates from a slight azimuthal alignment shift due to the sample temperature change.}
\label{fig:H2adsorption}
\end{figure}
Angular HAS scans were performed for the clean Sb(111) surface as well as after hydrogen exposure. Figure \ref{fig:H2adsorption} displays the resulting data along the $\overline{\Gamma \mathrm{M}}$ direction. Hydrogen adsorption was initiated by flooding the measurement chamber up to a total pressure of $1.2 \cdot 10^{-7}$~mbar for 4 minutes (22 Langmuir) as well as for 15 minutes (81 Langmuir) with the sample at room temperature. As can be seen in figure \ref{fig:H2adsorption}, neither the scattering intensities nor the peak positions changed significantly, indicating the total lack of adsorbed hydrogen on the antimony surface. The same holds when the surface is cooled down to 110 K and exposed to H$_2$ (lower panel in figure \ref{fig:H2adsorption}). Hence we can exclude the adsorption of molecular hydrogen at temperatures above 110 K for timescales relevant in single atom scattering experiments. Shorter sticking timescales for hydrogen on Sb(111) can be probed by directly scattering hydrogen molecules from the clean Sb(111) surface as has been done on several other metal surfaces\cite{Farias1998}. The Sb(111) surface seems to be largely inert with respect to molecular hydrogen adsorption - as has been indicated by J. Shan et al.\cite{Shan2011} - and hydrogen needs to be pre-dissociated before adsorption on the Sb(111) surface. Furthermore, since dissociation does not occur on the Sb(111) surface, this means that the upper limit for the H-Sb binding energy must be half the $\mathrm{H}_2$ binding energy of $4.52~$eV.\\

\subsection{Atomic Hydrogen Adsorption}
\begin{figure}[htb]
\includegraphics[width=0.8\columnwidth]{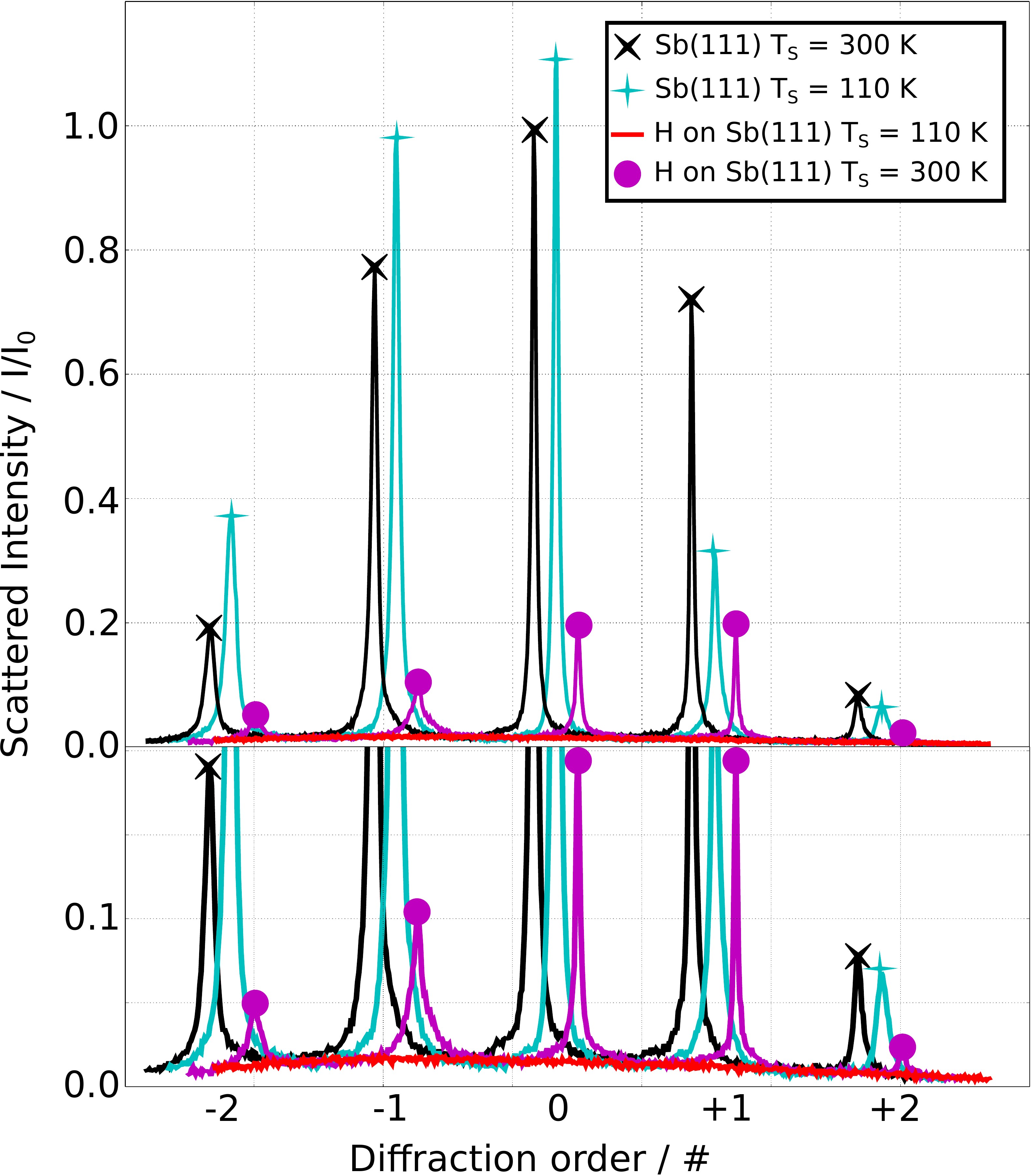}
\caption{Angular scattering spectra after several stages of atomic hydrogen exposure along the $\overline{\Gamma \mathrm{M}}$ direction of Sb(111). After dosing atomic hydrogen onto Sb(111) at 110~K the diffraction peaks are completely absent (red line), indicating a non-ordered overlayer. Upon heating the surface to 300~K, the diffraction pattern is recovered but with smaller different peak intensities (purple curve).}
\label{fig:Hadsorption}
\end{figure}

\begin{figure}[htb]
\includegraphics[width=0.8\columnwidth]{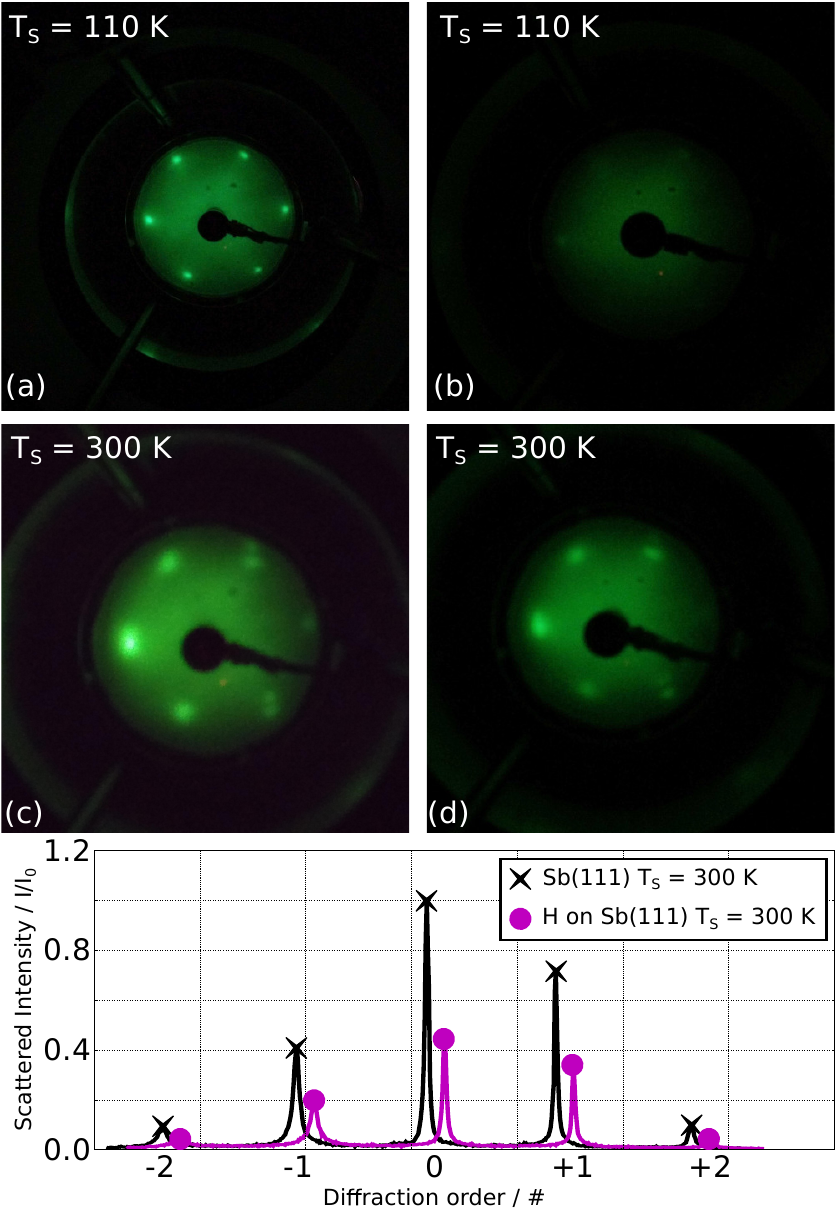}
\caption{Top: LEED images after dosing atomic hydrogen at two different surface temperatures. (a) Clean Sb(111) surface at $T_S = 110$~K; (b) Loss of structure after atomic hydrogen exposure at $T_S = 110$~K; (c) Clean Sb(111) surface at $T_S = 300$~K; (d) Almost no change in the LEED pattern, aside from a slight intensity loss after atomic hydrogen exposure at $T_S = 300$~K.\newline Bottom: After dosing atomic hydrogen onto the surface at $T_S = 300$~K, the resulting angular HAS scan reproduces the same pattern as observed after heating the hydrogen covered surface up to room temperature (see figure \ref{fig:Hadsorption}). This indicates the organisation of the hydrogen overlayer into an ordered structure.}
\label{fig:LEED}
\end{figure}

To study the effect of atomic hydrogen adsorption on the Sb(111) surface, the H$_2$ was thermally dissociated by heating the tungsten nozzle of the hydrogen source up to $1450^{\circ}\mathrm{C}$ using electron bombardment. For a first analysis - starting with a clean sample - angular reference scans were performed at 300~K and 110~K. A large amount of atomic hydrogen ($>1000$~L) was then dosed onto the cooled surface and immediately analysed by another angular scattering scan. The thus obtained surface was then heated up to 300~K and again monitored by an angular scan.\\
The scans are depicted in figure \ref{fig:Hadsorption}. After treating the cold surface with atomic hydrogen, the characteristic Sb(111) diffraction peaks are completely absent and only the diffuse multi-phonon background remains. This indicates a totally unordered, amorphous surface structure. Most likely the hydrogen atoms stick at the site where they arrive on the surface and the formation of any ordered structure is kinetically hindered.\\
Upon re-heating the surface to 300~K, distinct diffraction peaks re-appear, with lower intensity but in the very same spots as the Sb(111) peaks. This can either indicate the partial desorption of adsorbed hydrogen, leaving parts of the clean Sb(111) surface behind, or the formation of an ordered $(1\times 1)$ H overlayer. Since we obtain the same diffraction peaks upon heating the deposited H-layer (figure \ref{fig:Hadsorption}) and dosing atomic H with the crystal held at room-temperature (lower panel of figure \ref{fig:LEED}), the formation of an ordered layer of hydrogen seems more likely. Furthermore, LEED measurements presented in the upper panel of figure \ref{fig:LEED} support the interpretation of this behaviour.\\
The diffraction peak intensities of the ordered H layer and the clean Sb(111) surface are within the same order of magnitude. However, it is difficult to provide a quantitative comparison of the electronic corrugation without calculations of the diffraction peak intensities. The angular diffraction spectra lack pronounced resonance features in the inelastic background between the elastic scattering peaks. Such features, as observed on the clean Sb(111) surface previously \cite{Mayrhofer2013b} could be used to determine the atom-surface interaction potential and allow a detailed corrugation analysis using a close-coupling approach. Note, that the electronic corrugation of the clean Sb(111) surface ($0.6$ \AA\cite{Kraus2014}) is already larger than the corrugation of most hydrogen covered metal surfaces\cite{Lee1983,Gross1991}\\

\subsection{Thermal organisation and desorption}
\begin{figure}[htb]
\includegraphics[width=0.8\columnwidth]{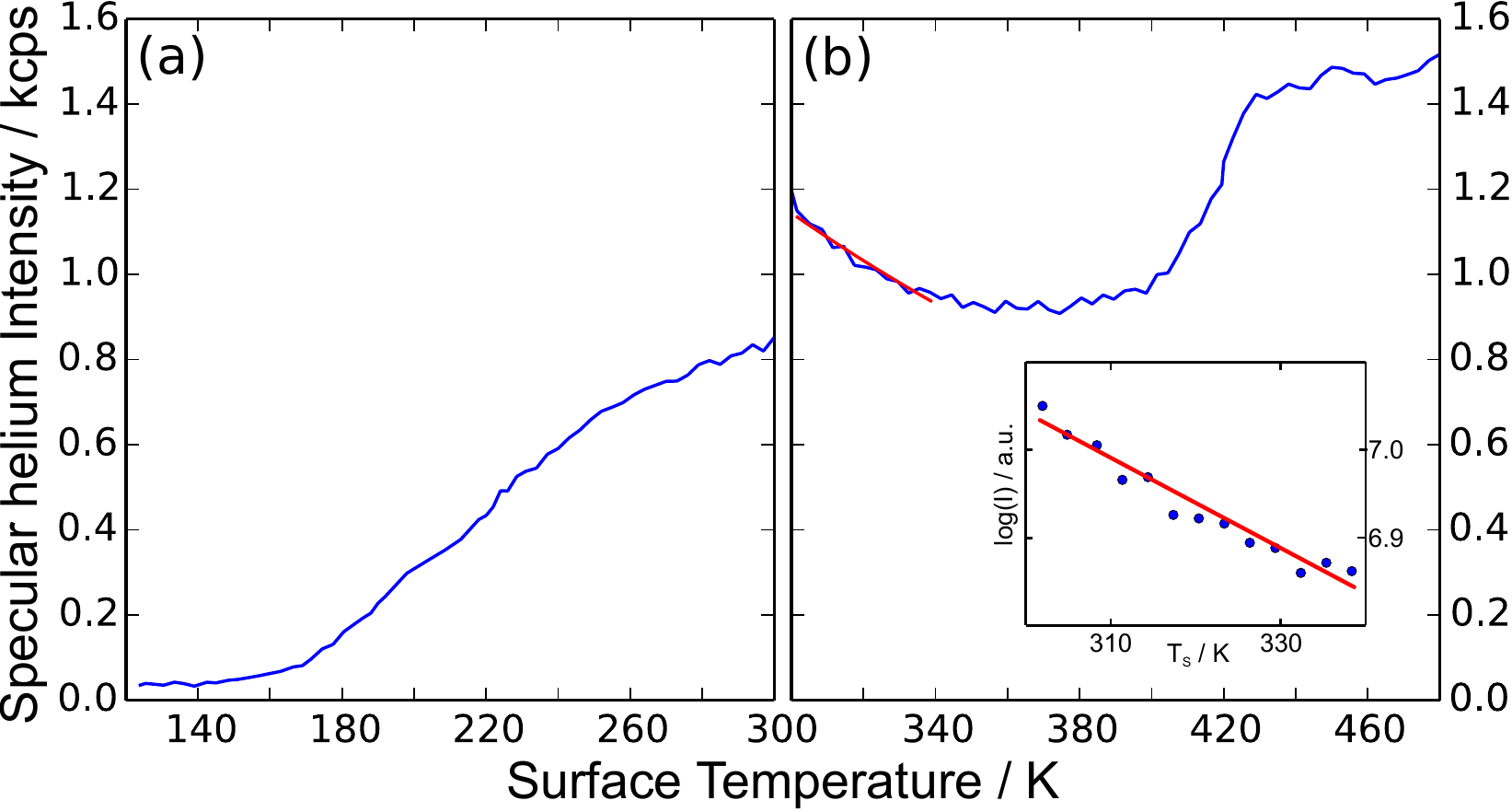}
\caption{Intensity of the specular helium reflection from a hydrogen covered Sb(111) crystal versus surface temperature. The starting point represents an amorphous hydrogen layer which had been deposited prior to the measurement. (a) At a surface temperature of $\approx150$~K, the signal slowly starts its continuous rise, indicating the start of an ordering process among the hydrogen atoms. (b) Further heating of the fully ordered hydrogen overlayer gives rise to a decline due to the Debye-Waller attenuation which is followed by a sudden increase in intensity at around $430$~K. This sudden change indicates the desorption of the hydrogen layer.}
\label{fig:H_drift}
\end{figure}
To determine the stability of the amorphous hydrogen layer as well as the binding energy of the crystalline hydrogen layer, the specular helium scattering intensity was recorded during the heating process. The surface under observation was sputtered and annealed before large amounts of atomic hydrogen ($>1000$~L) were dosed onto the cold surface at $T_S = 110$~K. As the machine is usually not prepared for conducting thermal desorption spectroscopy (TDS) experiments, the heating rate available was limited to $0.035~\mbox{K}\cdot\mbox{s}^{-1}$. Figure \ref{fig:H_drift} presents the two temperature drift measurements performed. Already after a small increase of the sample temperature, a continuous rise in the specular reflection commences. This dynamic enhancement of helium reflectivity reaches its maximum rate at around $250~$K and slows down significantly before reaching room temperature (Fig. \ref{fig:H_drift}(a)). The initial rise represents the onset for the transition from a disordered hydrogen overlayer to an ordered structure which has been characterised in the previous section. Once the diffusion of hydrogen on the surface is no longer kinetically hindered, the hydrogen atoms are able to reach their preferred adsorption sites until at 300 K the transition is almost complete.\\
In order to allow the overlayer to form the fully ordered structure, the system was kept at constant conditions for 8 hours. Upon further heating of the ordered hydrogen overlayer at $300~$K (Fig. \ref{fig:H_drift}(b)), the specular signal first drops according to the Debye-Waller attenuation as the surface temperature is raised, but exhibits a sharp rise around $430~$K. Since we expect that the hydrogen overlayer is in a completely ordered state by now, the feature can only be explained by thermal desorption of the hydrogen atoms, leaving the clean Sb(111) surface behind. Using this transition temperature together with the Redhead concept for first order desorption\cite{Redhead1962}, the desorption energy $E_{des}$ can be estimated. Redhead's equation reads as follows:
\begin{equation}
 E_{des} ~=~ k_B \cdot T_m \left[ \ln \left( \frac{\nu \cdot T_m}{\beta} \right) ~-~ 3.46 \right],
\end{equation}
with $T_m$ the temperature of the desorption maximum, $k_B$ the Boltzmann constant, $\beta$ the heating rate and the preexponential or frequency factor $\nu$.
Using $T_m = 430~$K as measured in figure \ref{fig:H_drift} and $\nu = 10^{13}$~s$^{-1}$\cite{Christmann1988} we obtain a desorption energy of about $(1.33\pm0.06)~$eV.\\
The inset in Fig. \ref{fig:H_drift} shows the logarithm of the specularly reflected intensity as a function of $T_S$ in the range 300-350 K for the
$(1\times1)$-H saturated surface. Even though only a limited temperature window is available the slope can be used to determine an estimate of the  Debye temperature\cite{Farias1998}. Hence the Debye temperature of the H-covered surface is $\theta_D = (530 \pm 20)~\mathrm{K}$ compared to  $\theta_D = (155 \pm 3)~\mathrm{K}$ for the pristine Sb(111) surface\cite{Tamtoegl2013}. This indicates the different vibrational properties of the hydrogen covered surface and is in accordance with other studies, e.g. the deuterium or hydrogen covered Ru(0001) surface, where the Debye temperature of the adsorbate covered surface is even increased by a factor of 5 and 8, respectively\cite{McIntosh2013}.\\

\subsection{Atomic hydrogen uptake}
\begin{figure}[htb]
\includegraphics[width=0.98\columnwidth]{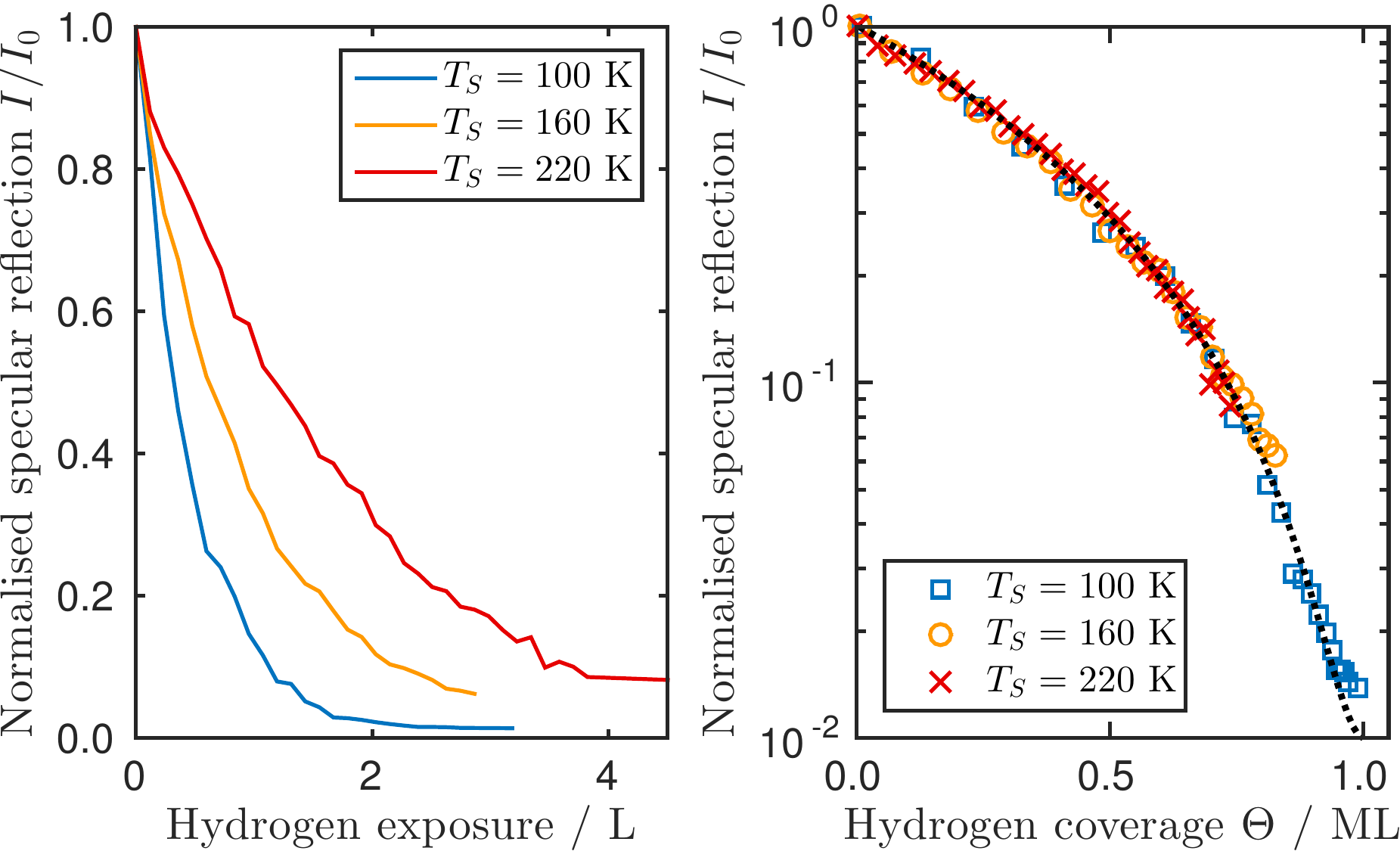}
\caption{Left panel: Normalised specular reflection $I/I_0$ versus exposure for the adsorption of atomic hydrogen on Sb(111) at three different surface temperatures and an incident beam energy of $16.2$ meV.\newline Right panel: $I/I_0$ versus coverage plotted on a logarithmic scale. The coverage has been determined from the exposure using the sticking coefficients listed in table \ref{tab:StickingCoeff}. The dashed line corresponds to equation \ref{eq:I0atten1} with $\Sigma = 12~\mbox{\AA}^{2}$.}
\label{fig:H_uptake}
\end{figure}

In order to investigate the adsorption of hydrogen on Sb(111), the helium specular signal $I$ was measured while dosing atomic hydrogen for several different surface temperatures $T_S$. The left panel in figure \ref{fig:H_uptake} shows the relative He specular peak height $I / I_0$ as a function of H exposure during adsorption for $T_S =100$, 160 and 220 K. The different slopes at the beginning clearly point to differences in the adsorption behaviour at these three temperatures. Furthermore, we notice that the curve at 100 K decays to almost zero whereas at higher temperatures, in particular at 220 K, $I/I_0$ levels off at a constant value greater than zero. Subsequent diffraction measurements after the dosing had stopped show, that the intensity remains constant provided the surface is held at a constant temperature. The behaviour at 100 K is typical if no ordered structure forms whereas at higher surface temperatures an ordered hydrogen overlayer forms which confirms the result  of the diffraction measurements in the previous section.\\
The fact that the presence of hydrogen on the surface substantially attenuates the specular beam indicates that the hydrogen covered parts of the surface do not contribute to the specular intensity. Hence the attenuation of the specular intensity during the adsorption of hydrogen can be used as a direct measure of the probability of diffuse scattering\cite{Poelsema1989}. In other words, the adsorbed hydrogen atoms scatter the He beam diffusively and the specular intensity arises exclusively from substrate areas not covered by hydrogen. The normalised specular intensity $I / I_0 $ can then be related to the hydrogen coverage $\Theta$ via:
\begin{align}
I / I_0 &= \left( 1 - \Theta  \right) ^ {n \cdot \Sigma / \cos \vartheta_i } \label{eq:I0atten1} \\
& \approx 1  - \Theta \cdot n \cdot  \Sigma / \cos \vartheta_i \quad \mbox{for} \quad \Theta \ll 1
\label{eq:I0atten2}
\end{align}
where $n$ is the adsorbate density at monolayer (ML) coverage, $\Sigma$ is the helium scattering cross section and the term $\cos (\vartheta_i )$ accounts for the increase of the apparent scattering cross section since scattering happens at an incident angle $\vartheta_i = 45.7^{\circ}$. Equation \ref{eq:I0atten1} follows a strict geometrical overlap approach of the scattering cross sections and assumes no interaction between the adsorbates (random adsorption of the adsorbates)\cite{Poelsema1989,Farias1998}.\\
In case of low coverage and no adsorbate-adsorbate interaction, a linear dependence of the intensity on the coverage can be assumed (equation \ref{eq:I0atten2}) and the scattering cross section $\Sigma$ can be determined from the initial slope of the adsorption curve\cite{Farias1998}. Hence the different slopes in figure \ref{fig:H_uptake} suggest that either the helium scattering cross section changes with temperature or the sticking coefficient decreases with temperature. Since it is unlikely that $\Sigma$ changes with surface temperature we assume that the sticking coefficient becomes smaller with increasing surface temperature.\\
The right panel in figure \ref{fig:H_uptake} shows $I/I_0$ versus coverage where the surface coverage $\Theta$ has been determined from the exposure using an initial sticking coefficient of 1 for adsorption at 100 K. Note that this assumption can only serve as an upper bound and is based on the fact that the sticking coefficient for atomic hydrogen on metal surfaces is often close to one\cite{Winkler1998}.\\
Monolayer coverage consists of one hydrogen atom per unit cell (as also confirmed by the diffraction data in the previous section) which corresponds to an adsorbate density $n=0.108~\mbox{\AA}^{-2}$. The surface coverage is related to the surface exposure which is defined as the impinging flux of adsorbates on the surface integrated over the time of exposure. For this purpose, the dosing pressure was corrected according to\cite{Winkler1998}, to obtain the number of hydrogen atoms impinging on the surface compared to the pressure readout. Furthermore we have used a Langmuir adsorption model where adsorption is limited to one monolayer, all adsorption sites are equivalent and only one particle can reside in an adsorption site. Therefore the sticking coefficient is given by $S(\Theta)=(1-\Theta ) S_0$ where $S_0$ is the initial sticking coefficient for the uncovered surface.\\
In order to maintain a constant helium scattering cross section, $S_0$ needs to decrease with increasing temperature. The results for $S_0$ which give rise to the same initial slope are given in table \ref{tab:StickingCoeff}. As can be seen from the right panel in figure \ref{fig:H_uptake} the hereby determined adsorption curves at different surface temperatures are now on top of each other.\\
\begin{table}
\centering
\caption{Temperature dependence of the initial sticking coefficient for the adsorption of atomic hydrogen on Sb(111) relative to the sticking coefficient at 100 K.}
\begin{tabular}{ l | r r r}
\midrule
Surface Temperature $T_S$ (K) & 100 & 160 & 220 \\ 
\midrule
Initial sticking coefficient & \multirow{ 2}{*}{1.0} & \multirow{ 2}{*}{0.54} & \multirow{ 2}{*}{0.32} \\ 
$S_0 (T_S ) / S_0(T_S = 100~\mathrm{K})$ \\
\midrule
\end{tabular}
\label{tab:StickingCoeff}
\end{table}
Finally, the dashed line in figure \ref{fig:H_uptake} corresponds to a fit according to equation \ref{eq:I0atten1} where we have added a small offset at monolayer coverage. The best fit value of the helium scattering cross section $\Sigma$ is $(12 \pm 1)~\mbox{\AA}^{2}$. While for hydrogen on Fe(110) $\Sigma = 3.5~\mbox{\AA}^{2}$ (for 63 meV beam energy\cite{Farias1998}), the scattering cross section for hydrogen on transition metal surfaces as well as on metal oxide surfaces is typically around $10-12~\mbox{\AA}^{2}$ (for 11-40 meV beam energy\cite{Poelsema1989,Woell2004}), which is in good agreement with our result.

\section{Conclusion}
\label{sec:conclusion}
We have carried out a series of helium atom scattering experiments in order to characterise the adsorption properties of hydrogen on Sb(111).
Initial HAS spectra show that molecular hydrogen does not spontaneously dissociate and adsorb on the Sb(111) surface. The astonishing inertness of the Sb(111) surface towards the adsorption of molecular hydrogen explains why the Sb(111) surface remains exceptionally clean in UHV environments where the main residual gas contribution is molecular hydrogen.\\
Depending on the substrate temperature while dosing pre-dissociated atomic hydrogen onto the antimony surface, two different adlayer phases were observed. At low substrate temperatures (experiments performed at $T_S = 110~$K), the resulting hydrogen overlayer observed by HAS and LEED does not show any ordering. The same experiments performed with the surface at room temperature revealed a perfectly ordered $(1\times 1)$ H/Sb(111) structure. Furthermore, the hydrogen layer deposited at low temperature appears to re-organize upon heating the crystal to room temperature.\\
When monitoring the specular He reflection upon heating of the hydrogen covered surface, hydrogen starts to desorb at $T_m = 430~$K which corresponds to a desorption energy of $E_{des}=(1.33\pm0.06)~$eV. The attenuation of the specular He reflection during the adsorption of atomic hydrogen can also be used as a measure of the hydrogen coverage. At low surface temperature (100 K), the specular intensity is completely attenuated after large exposures whereas at higher surface temperatures, it levels off at a constant value. This is another proof for the formation of an amorphous overlayer at low temperatures and an ordered structure at higher crystal temperatures.\\
Moreover, from the different slopes of the adsorption curves we conclude that the initial sticking coefficient of atomic hydrogen on Sb(111) decreases with increasing surface temperature. The scattering cross section for the diffuse scattering of helium from hydrogen on Sb(111) is determined with $\Sigma = (12 \pm 1)~\mbox{\AA}^{2}$, which is in good agreement for hydrogen on other surfaces.

\acknowledgments
We would like to thank A. Winkler from the Institute of Solid State Physics (Graz University of Technology) for providing us with the atomic hydrogen source and for his advice on molecular hydrogen dissociation. One of us (A.T.) acknowledges financial support provided by the FWF (Austrian Science Fund) within the project J3479-N20.

\bibliographystyle{unsrt}
\bibliography{HSb_literatur}

\end{document}